\documentclass[ aps, reprint, amsmath, amssymb ]{revtex4-1}
\usepackage{graphicx}
\usepackage{dcolumn}% Align table columns on decimal point
\usepackage{bm}% bold math
\usepackage[usenames]{color}
\usepackage{colortbl}

\begin {document}
%\global\textheight=239mm
%global\headsep=0.1mm
\title{Universal cluster size distribution in a system of randomly spaced particles}
\author{Khokonov M.Kh.}
 \altaffiliation{Kabardino-Balkarian State University, Nalchik, Russian Federation.}
 \email{Electronic address: khokon6@mail.ru}     

\author{Khokonov A.Kh.}
\altaffiliation{ Kabardino-Balkarian State University, Nalchik, Russian Federation.}
 \email{Electronic address: azkh@mail.ru}
\date{\today}

\begin{abstract}
The distribution function of particles over clusters 
%consisting of them - Changed!! 
is proposed for a system of identical intersecting spheres, the centers of which are uniformly distributed in space.
%\textcolor{blue}{
Consideration is based on the concept of  the rank number of clusters, where the rank is assigned to clusters according to the cluster sizes. 
%}
Distribution is universal in the sense that it 
%\textcolor{blue}{
does not depend on boundary conditions  and
%}
is valid for infinite medium. The form of the distribution function is determined by only one parameter, equal to the ratio of the sphere radius (‘interaction radius’) to the average distance between the centers of the spheres. This parameter  plays also a role of the order parameter.
It is revealed under what conditions the universal distribution
%\textcolor{blue}{
behaves like well known   log-normal distribution.
%} 
Applications of the proposed distribution to some realistic physical situations, which are close to the conditions of the  
%\textcolor{blue}{
gas condensation to liquid,
%} 
are considered. 
\end{abstract}

\maketitle

%\vskip1cm
%\thispagestyle{empty}

\section{Introduction}

%\noindent
%\textcolor{blue}{Checking a color}

Consider a system of identical point particles randomly distributed within a certain volume with a density $\rho=N_0 /V$, where 
$N_0 $ is  the total number of particles in the system with volume $V$.  The most probable average distance between particles in such a system is $l_0=0.55396\rho^{-1/3}$ (see Hertz distribution, Eq. (676) in \cite{Chandrasekhar}).  We assume that two particles belong to the same cluster, if the distance between them does not exceed a certain ``radius of interaction'' $R$.  The system, therefore, is a collection of identical spheres, some of which intersect with each other, forming clusters. We are interested in the question, what is the probability that a cluster contains exactly $N$  particles. 

%\textcolor{blue}{
Analytical solution of this problem has not been found at present. The question is interesting independently on its possible applications. We consider this problem numerically and express results in terms of some one-parameter distribution function which is universal in that sense that it does not depend on the boundary conditions (i.e. on the volume of the system for fixed particle density). If the numerical values of this function are tabulated, then one can calculate all the quantities  related to the task at hand which are of practical interest.
% }

In our model, we assume that particles can move closer to any distance without affecting each other. Intersecting spheres form the ``bound states''. Then, as it will be seen below, the distribution of particles over clusters will be determined by only one parameter
% -------------------------- i ---- (1) -------
\begin{equation}
a=R/l_0.
\label{a}
\end{equation}
 We shall call the quantity $a$ ``the interaction parameter''. This parameter defines the system configuration and plays for it the role of the order parameter.

%\textcolor{blue}{
The idea that the geometry of a microstructure contains an important physical information of the constituent solids and liquids was put forth by Smith more than seven decades ago \cite{Smith1948}. 
The statement of our problem shows that the centers of the spheres are distributed according to Poisson's law. A large body of work has been performed on spheres centered according to such distribution \cite{Meijering1953},  \cite{Moller1989}. 
More recent applications of this to granular materials and Lorentz gasses are discussed in detail in \cite{Morse2014}, \cite{Yuliang2015}.  
In contrast to these works, the clusters in our case are not necessarily represent three-dimensional convex structures, but can have an arbitrary shape.
%}

The cluster size distribution is of primary importance in physics of aerosols \cite{Vincent2007}, powder, or granular materials \cite{deBono2020}, \cite{Allen2003},  cataclastic fault materials \cite{LinJi1994},
 as well as the mass distribution among the recovered fragments of the meteorites 
\cite{Gritsevich2014}. The  continuous type distributions, among other things, describe the mass-size distribution of aeolian sand deposits and are of some potential usefulness in other concrete contexts too \cite{Barndorff1977}, \cite{Sorensen2016}. 
%\textcolor{blue}{
The model of overlapping spherical particles was used in \cite{McDowell2013} to determine the size distribution of 
fragmented particles under normal one-dimensional compression.
%}

The size distributions are rather complex as they can not be fitted into a single function \cite{Nguyen2014}. Most of the distribution functions used in practice are phenomenological, with the exception of some of them having a mathematical basis. 
Kolmogorov suggested that the probability of fragmentation does not depend on the size  during 
the crushing process, and showed that the spectrum of grain sizes tends to the log-normal distribution \cite{Kolmogorov1941}.

It has recently been shown that distribution of the contact forces in granular materials during confined comminution follow 
a clear log-normal distribution as well \cite{Ben-Nun2010}. Significant progress has been made 
in the consideration of the properties of non-spherical clusters  \cite{Azema, *Azema1, *Azema2, *Azema3}.
This case is covered by the present approach, since the form of clusters in a system of 
randomly located spheres can be arbitrary. We emphasize that we consider static systems. However, 
we will discuss also the extent to which the results obtained can be applied to dynamic systems as well.

%\cite{Azema}
%\cite{*[{
%E. Az\'ema, and F. Radja,  Phys. Rev. E  {\bf 85}, 031303 (2012);
%E. Az\'ema, F. Radja, and  F. Dubois,  Phys. Rev. E {\bf 87}, 062203 (2013).
%}] Azema1}.
%E. Az\'ema, and F. Radja,  Phys. Rev. E  {\bf 81}, 051304 (2010); 

In what follows we shall consider a simple off-lattice  model which is more related to the conditions of condensation, and which, 
%\textcolor{blue}{
under some conditions,  reduces to the behaviour like a 
%}
 log-normal distribution. In contrast with conventional description based on the distribution of clusters over their sizes we study the distribution of particles among the clusters.
% that consist of them. Deleted !!

We do not address the problems of percolation \cite{Stauffer}, \cite{Hunt}, which are closely linked to the problem of cluster formation. Percolation assumes the existence of some boundary conditions. We study the distribution of particles over clusters independently on their shape in the infinite medium rather than the distribution of connected clusters in a random graph. 
The mathematical formulation of the problem noted above is to a certain extent similar to the percolation problem for semi-permeable spheres \cite{Rottereau2003}, \cite{Johner2008}. 

%\textcolor{blue}{
We also do not concern the geometric  size of the clusters. Instead, we study the numbers of particles in them. Clusters with the same size may have different numbers of particles, since the spheres representing particles may overlap. In this connection we do not concern the questions of packing fraction or volume ratio of spheres to the system volume. The latter can be infinite. These problems would be of interest for spheres with a solid core of a smaller radius, so that the centres of the spheres could not approach each other at a distance smaller than the size of the solid core \cite{Rottereau2003}, \cite{Johner2008}.  In this paper, the size of a cluster does mean the number of particles in it, and not the geometric volume.
%}

\section{Discrete distribution functions}

Let us renumber the clusters in decreasing order of the number of particles of which they are composed.
%We number the clusters in decreasing order of the number of particles in each cluster. 
That is, by definition, $k = 1$,  is the sequence number of the largest cluster containing the maximum number of particles $ N_1 \equiv N_{max}$. 
In this case, the numbers $k$  are not just the sequence numbers of clusters, but the numbers of clusters arranged in a certain order, therefore the numbers $k$  will also be called the ranked number  of the cluster.

Let $N_k$  be the number of particles in the cluster with a ranked number $k$. Obviously, by the definition of these numbers,  always $N_k \ge N_{k+1}$   and
% -------------------------- i ---- (2) -------
\begin{equation}
\sum_{k=1}^{N_0} N_k= N_0 .
\label{sum1}
\end{equation}
If the maximum value of $k$ for which $N_k \ne 0$  is $k_0$, then the summation in this formula can be extended to $k_0$.  At the same time, the number $k_0$  determines the number of clusters in the system, including clusters containing only one particle.

\begin{figure}
\centering
\includegraphics[bb=50 0 618 400, width=0.55\textwidth]{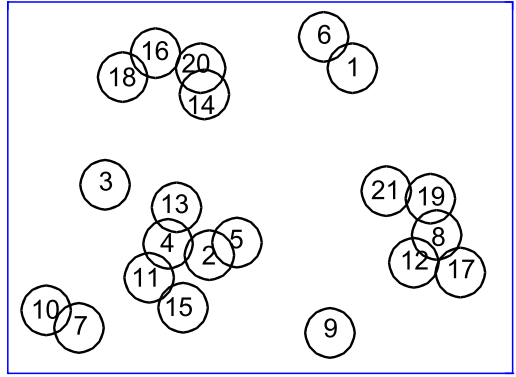}
% original eps: bb=0 0 568 397
\centering
\caption{
Clusters in a system of $N_0=21$  particles 
}
\label{fig_N=21}
\end{figure}

\begin{figure}
\includegraphics[scale=0.85]{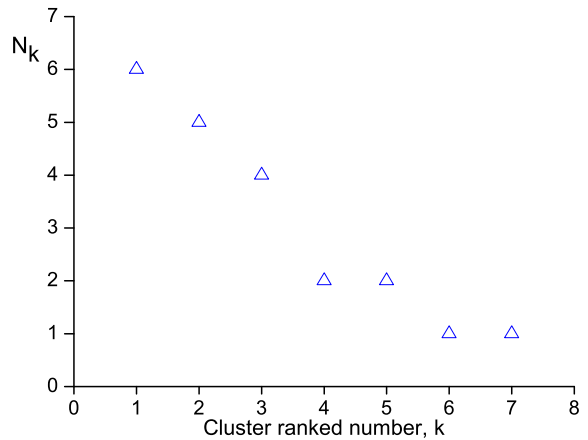}
%\includegraphics[bb=50 0 620 400, width=1.1\textwidth]{N_21_N_k.eps}
% original eps: bb=0 0 570 400
\centering
%\vspace{10mm}
\caption{
The distribution of particles over clusters $N_k$  in the system shown in Fig. \ref{fig_N=21}.
 }
\label{fig_N=21a}
\end{figure}

Fig. \ref{fig_N=21}   shows a system of $ N_0=21$  particles that are randomly distributed in a two-dimensional space. For this system, we have $ N_1=6$, $ N_2=5$, $ N_3=4$, $ N_4=N_5=2$, $ N_6=N_7=1$.  Fig. \ref{fig_N=21a}  illustrates the distribution of particles over clusters in this system $ N_k$ as a function of the ranked number $k$.
% ----------------------------------------------------------------------------------------------------------------------------
% ---------------------------------------------- fig 1------------------------------------------------------------------------------
% ----------------------------------------------------------------------------------------------------------------------------
% ----------------------------------------------------------------------------------------------------------------------------
% ---------------------------------------------- fig 2------------------------------------------------------------------------------
% ----------------------------------------------------------------------------------------------------------------------------
% ---------------------------------------------- ------------------------------------------------------------------------------
% ----------------------------------------------------------------------------------------------------------------------------

%Traditionally, the distribution function of clusters by size is used in practice. 
The quantity traditionally used in practice is the distribution function of clusters by size. 
In this connection we also introduce a discrete distribution function $W(N)$, which determines the 
probability that  cluster contains exactly $N$  particles,  that is  the distribution of clusters over the number of particles. We shall mainly  be interested, however, in the distribution of particles over clusters $N_k$. 
It will be seen from the following that  the distribution $N_k$   is  convenient in practice, 
especially under conditions close to the condensation. Condensation does mean such   configuration of a system when most particles are contained in several large clusters. The  distribution $N_k$  determines how the particles are distributed over clusters, while $W(N)$ determines the distribution of clusters over the number of particles. We take the distribution $N_k$ as a basis.
The advantages of the function $N_k$  are related to the fact that if, for example, the system is close to the conditions of condensation, then it may happen that the main fraction of the atoms of a substance is concentrated in a relatively small number of large clusters. Then the values of $W(N)$  for such particles may be very small, and the shape of this function will be determined by a large number of clusters with a very small number of particles, but which do not determine the characteristic properties of the system. On the contrary, the distribution $N_k$  distinguishes the largest clusters, which makes this distribution the most preferred 
%\textcolor{blue}{
in some practical applications.
%}

The analogue of the distribution function $W(N)$, which is generalized to continuous values of the argument, is traditionally used in physics of aerosols, powder, or granular materials.
% and is determined by distributions of the Kolmogorov type, log-normal distribution, H.Yung distribution, Hrgian-Mazin distribution, etc. \cite{}. 
At this stage we will focus on clarifying properties of discrete distributions normalized to the number of particles (or clusters) in the system.

Distribution functions $N_k$  and $W(N)$  are uniquely related.  Consider a discrete function $k(N)$, which is the inverse function of $N_k \equiv N(k)$. Then the quantities
 % -------------------------- i ---- (3) -------
\begin{equation}
W(N)=k(N) - k(N+1),
\label{WN}
\end{equation}
will determine the distribution $W(N)$, that is, the number of clusters containing exactly $N$   particles, where $1\le N \le N_{max} =N_1$, here $N_{max}$  is the number of particles in the largest cluster.
%\textcolor{blue}{
The integer-valued function $k(N)$  can be obtained by a $90^{\circ}$  clockwise rotation of the function $N_k$ 
followed by a reflection relative the abscissa axis. This new abscissa axis shows the number of particles $ N$, 
and the ordinate axis shows values of $k(N)$.
%}

Eq.  (\ref{WN}) can be understood from the following reasoning. The value of $k(1)$  is equal to the number of clusters containing at least one particle, while $k(2)$  gives  the number of clusters containing at least two particles. Then the difference between them will give the number of clusters containing exactly one particle, i.e. $W(N=1)$. It should be set  $k(N_{max} +1)= 0$  in Eq. (\ref{WN}).
%\textcolor{blue}{
According to the definition, $N_k$  represents a number of particles in a cluster with a rank number $k$. If clusters with numbers of particles $N_k-1, N_k-2, …, N_k-m$  are absent, then the same rank number $k$  must be taken in Eq. (\ref{WN}) to all these $m$ clusters. 
%}

 It is obvious that following relations take place
 % -------------------------- i ---- (4), (5) -------
\begin{eqnarray}
\sum_{N=1}^{N_{max}}  N W(N)=N_0, 
\label{sum2} \\
\sum_{N=1}^{N_{max}}   W(N)=k_0 . 
\label{sum3}
\end{eqnarray}
Thus, if the distribution $N_k$  is normalized to the number of particles in the system, then the distribution function $W(N)$  is normalized to the number of clusters. In the example with $N_0 = 21$, shown in Figs. \ref{fig_N=21} and \ref{fig_N=21a}, the function $k(N)$ takes 6 nonzero values:  $ k(1) = 7$,  $ k(2) = 5$, $ k(3) = k(4)=3$, $ k(5) = 2$ , $ k(6) = 1$ . Then the values of the distribution function $W(N)$  are as follows: $W(1) = W(2)=2$, $W(3) =0$, $W(4) = W(5)=W(6)=1$. In this example, $ N_{max} = 6$   and $k_0 = 7$. It is easy to verify that relations (\ref{sum2}) and (\ref{sum3}) are satisfied.

\section{Monte-Carlo simulation}

The distribution function $N_ k$ was calculated by the Monte-Carlo method. Each run for a fixed value of the parameter $a$ was carried out as follows. The system is a cubic volume of an arbitrary value $V = L^ 3$,  where $L$ is the length of the cube edge. This volume contains $N_0$ particles. For given $a$ and $N_0$, the relation between $R$ and $L$ is, $R = a l_0 =0.55396 a  L N_0^{-1/3}$. The numerical value of the dimensionless quantity $L$ does not affect the results of calculations and is determined by considerations of convenience (this is tantamount to the choice of the units of length). We took $L$ proportional to $N_0^{1/3}$. For $N_0 = 10^4$ we took $ L = 100$.  We could also change the definition of $l_0$, and take, for example, $l_0=\rho^{-1/3}$. In this case, all the results will remain unchanged, taking into account the scale redefinition of $a$.

The coordinates of centers of spheres with radius $R$ were simulated by means of three random numbers, uniformly distributed in the interval $[0, 1]$. For the configuration obtained in this way, the distribution of clusters $N_k$  was calculated, similar to the procedure shown in Fig.2,  but in three dimensions. The algorithm for such a calculation is given in the Appendix. 
Our algorithm is different from the cluster multiple labeling technique used in the  percolation theory \cite{Hoshen1976}. We are interested in all particles and clusters of the system, regardless of the percolation conditions. 

We performed $M=50$ runs for fixed $a$ and $N_0$. Each $i$-th run gives its own distribution $N_k^{(i)}$ and we may calculate the mean number of particles in a cluster with a rank number $k$, $\langle N_k \rangle$,  and the root mean square, 
$\langle \Delta N_k^2 \rangle^{1/2}$, where $\langle (...) \rangle = \sum_{i=1}^{M} (...)^{(i)}/M$, and 
$\langle \Delta N_k^2 \rangle = \langle  N_k^2 \rangle - \langle  N_k \rangle^2$. 
Some of the figures below show single-run calculations. Such cases will be noted separately. 

The set of distributions $N_k^{(i)}$ permits  to calculate (for fixed $k$)  the probability that the cluster with
the ranked number $k$ contains exactly $n$ particles,  $w^{(k)}(n)$.  
The distribution function $w^{(k)}(n)$  is normalized to unity (see Eq. (\ref{sum_w})  below). It represents the solution of the problem outlined in the beginning of the Introduction, and also permits one to calculate the average values $\langle N_k \rangle$ 
(see  Eq. (\ref{sumk_w})).

The single-run function $N_k$  leads to the corresponding integer-valued probability $W(N)$,  which shows the number of clusters with $N$  particles in agreement with Eq.(\ref{WN}). After averaging over all runs, functions $\langle N_k \rangle$ and $\langle W(N) \rangle$  are not integer-valued. In some cases, however, it is reasonable to round them off to the integer numbers, because any single measurement over the system results in integer function. If any single-run function $W^{(i)}$  is normalized to unity with its own normalization constant, the resulting averaged function will be normalized to unity as well.

In what follows we shall retain the notation $N_k \equiv \langle N_k \rangle$ (as well as, $W(N) \equiv \langle W(N) \rangle$) for the average numbers, which are plotted in the figures below. The results of such calculations will depend on two numbers, $a$ and $N_0$.  
The dependence on the number of particles in the system $N_0$  is due to the influence of the boundaries on the form of the distribution. This influence is stronger, as $N_0$  is smaller. With an increase in $N_0$  (recall that all these arguments refer to systems with a fixed value of the parameter $a$), the influence of the boundaries of the cube weakens, since the fraction of particles on its surface decreases in comparison with that  in the volume. It will be seen from what follows, that for large $N_0$, one can introduce some new distribution, $N(x)$, 
%\textcolor{blue}{
where x is a continuous variable, 
%}
$0 <x <1$, depending on $a$, but independent of $N_0$, that is, on the boundary conditions.

%The numbers $N_k$  were determined by means of the Monte-Carlo simulation for a spatially uniform distribution of particles in the system, for fixed total  number of particles  $N_0$  and for a given value of the interaction parameter $a$. The set of occupation numbers of clusters with particles $N_k$ is a random variable, and each such set has its own probability. For example, one can ask the question what is the probability that a cluster with ranked number $k$ contains exactly $n$ particles. We calculated the average values of the occupation numbers, as a result of averaging over 50 runs. These average values will also be denoted as $N_k$.

\section{Results of computer simulation}

% ----------------------------------------------------------------------------------------------------------------------------
% ---------------------------------------------- fig 3------------------------------------------------------------------------------
% ----------------------------------------------------------------------------------------------------------------------------
\begin{figure}
\includegraphics[bb=50 0 620 400, width=0.55\textwidth]{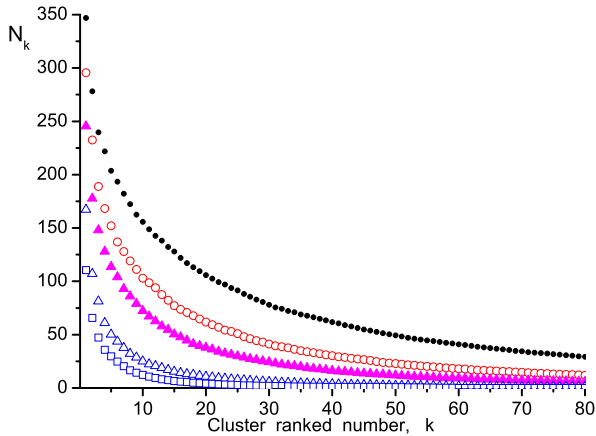} 
% original eps: bb=0 0 570 400
\centering
%\vspace{15mm}
\caption{
The distribution of particles over clusters $N_k$  for interaction parameter $a=2$, as a function of the ranked number  $k$  for different particle numbers $N_0$ in the system: dots -- $N_0=10^{4}$; circles -- $N_0=5000$; dark triangles  -- $N_0=3000$; open triangles -- $N_0=1000$; squares -- $N_0=500$ 
 }
\label{fig_a=2}
\end{figure}
% ---------------------------------------------- ------------------------------------------------------------------------------
% ----------------------------------------------------------------------------------------------------------------------------

The distribution of particles with $a=2$  over clusters $N_k$  as a function of the ranked cluster number $k$  for different values of the number of particles (atoms) in the system $N_0$   is shown in Fig.3. 
%\textcolor{blue}{
The results are averaged over 50 Monte-Carlo runs.
%}

It follows from Fig. \ref{fig_a=2} that in a system with 10,000 atoms there is one biggest cluster with an average number of particles in it  $N_1 \approx 350$; one cluster with $N_2 \approx 275$; clusters with numbers $k=$45-55 contain approximately 50 particles each. For a given number of particles in the system $N_0$, the distribution  $N_k$   is completely determined by the value of the interaction parameter $a$. It is clear that the greater the  $N_0$, the higher position of the curves in Fig. \ref{fig_a=2}.

% ----------------------------------------------------------------------------------------------------------------------------
% ---------------------------------------------- fig 4 ------------------------------------------------------------------------------
% ----------------------------------------------------------------------------------------------------------------------------
\begin{figure}
\includegraphics[bb=50 0 620 400, width=0.55\textwidth]{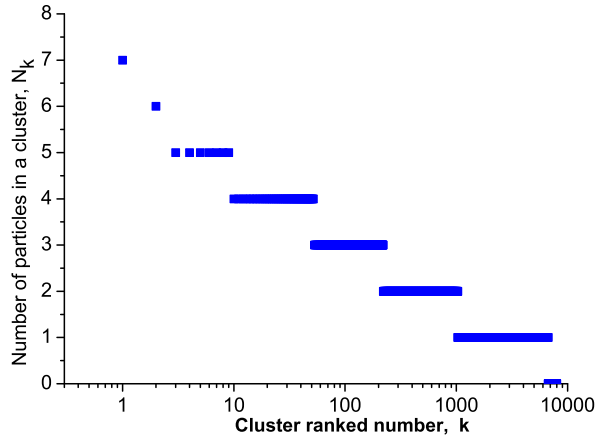} 
% original eps: bb=0 0 570 400
\centering
\caption{
Particle distribution over clusters $ N_k$  as a function of the ranked number $k$  with the value of the interaction parameter $a = 0.8$  in a system of $N_0$ = 8000 particles.
}
\label{fig_N_8000}
\end{figure}
% ----------------------------------------------------------------------------------------------------------------------------
% ---------------------------------------------- fig 5 5------------------------------------------------------------------------------
% ----------------------------------------------------------------------------------------------------------------------------
\begin{figure}
\includegraphics[width=0.55\textwidth]{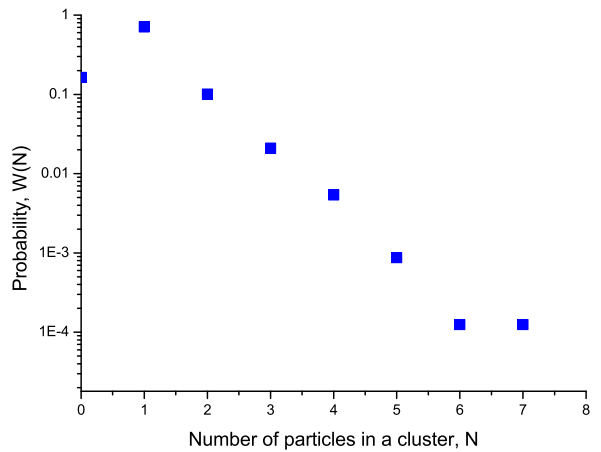}  
% original eps: bb=0 0 570 400
%\vspace{5mm}
\centering
\caption{
The distribution of clusters by the number of particles
% in themc- Deleted !! 
$W(N)$  for the system shown in Fig. \ref{fig_N_8000}.
}
\label{fig_W_8000}
\end{figure}
% ----------------------------------------------------------------------------------------------------------------------------
% ----------------------------------------------------------------------------------------------------------------------------

Figs. \ref{fig_N_8000} and \ref{fig_W_8000}  illustrate the difference in the distribution functions $N_k$  and $W(N)$  for the same system with $N_0$ = 8000 and  $a$ = 0.8.   
%\textcolor{blue}{
These figures 
%Figs. \ref{fig_N_8000} and \ref{fig_W_8000}  
represent a single-run calculations, where  a discrete function 
 $W(N)$  is normalized to unity.
%}
This is the case when the cluster formation has just begun and the largest cluster contains only 7 particles, although the total number of clusters is large, $k_0=6687$. $W(0)$  in Fig. \ref{fig_W_8000}  indicates the probability that a cluster does not contain particles. This number (in this example, $W(0) $ = 0.164)  means that the fraction  of particles that are included in at least one cluster (including clusters consisting of one particle) is equal to $k_0/N_0=0.836$.

% ----------------------------------------------------------------------------------------------------------------------------
% ---------------------------------------------- fig 6 6------------------------------------------------------------------------------
% ----------------------------------------------------------------------------------------------------------------------------
\begin{figure}
\includegraphics[width=0.475\textwidth]{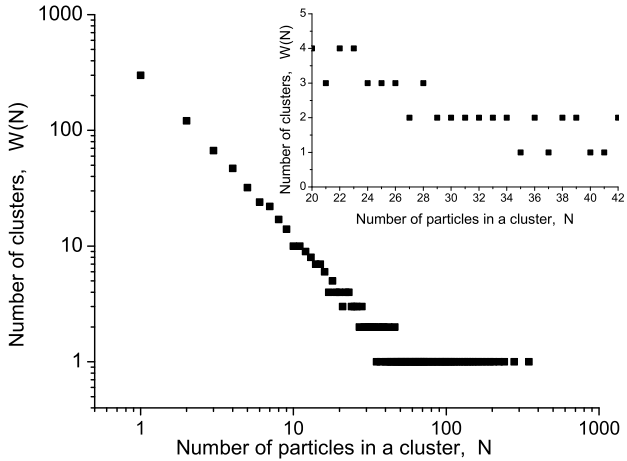}  
% original eps: bb=0 0 570 400
%\vspace{5mm}
\centering
\caption{
The distribution of clusters by the number of particles $W(N)$  for $a=2$ and $N_0=10^4$. $W(N)$  is normalized according to Eq. (\ref{fig_W_8000}). The upper plot shows the particle number interval, $20<N<42$,
 on an enlarged linear scale.
}
\label{fig_W_2_10000}
\end{figure}
% ----------------------------------------------------------------------------------------------------------------------------
% ----------------------------------------------------------------------------------------------------------------------------

%\textcolor{blue}{
The distribution of clusters, $W(N)$,  over the number of particles for larger value of the parameter $a=2$ is shown in Fig. \ref{fig_W_2_10000} for $N_0=10^4$. In contrast with   Fig. \ref{fig_W_8000}, the distribution $W(N)$ is obtained from the distribution $N_k$, averaged over 50 runs, and corresponds to the upper curve in Fig.  \ref{fig_a=2}. $W(N)$ on this figure is normalized as in Eq. (\ref{fig_W_8000}) (in this case, $k_0=824)$,   
and is rounded off to the integer numbers. It shows the number of clusters as a function of number of particles $N$.  The number (or fraction) of clusters in this case is much less than that shown in Fig.   \ref{fig_W_8000}  but clusters are bigger in size. The biggest cluster contains $N_1=347$ particles.     The function $N_k$  always decreases monotonically, while the function $W(N)$  does not behave monotonically and regularly, even for single-run calculation. This is clearly seen in the upper graph  in Fig. \ref{fig_W_2_10000},      
 although in general the function $W(N)$  decreases with increasing $N$.            
%}
 
Figs. \ref{fig_N_8000} and \ref{fig_W_8000}  show that 
the description of the system in terms of the  distribution over the number of particles $ N_k$  is
%\textcolor{blue}{
 more convenient  
%}
than the description using the distribution over the clusters  $W(N)$,
%\textcolor{blue}{
if we are interested in largest clusters in the system,  since such clusters correspond to the tail of the distribution $ W(N)$  with  a very small values of $ W \sim 10^{-4} $.
%}

%\textcolor{blue}{
Distribution $N_k$  does not fully define the statistical properties of the system. 
%}
As it was mentioned above, the values of $N_k$ are the average values of the numbers of particles in a cluster with the ranked number $k$. In fact, for a given $k$, there is a certain distribution of clusters over the number of particles in it. That is, for example, the number of particles in the largest cluster $N_1$ is a random variable, and the values of $N_1$ shown in Fig. \ref{fig_a=2} for different $N_0$ are average values of $N_1$. 

Let $w^{(k)} (n)$  be the probability that the cluster with a ranked number $k$ contains exactly $n$ particles, then for all $ k$
 % -------------------------- i ---- (6) -------
\begin{equation}
\sum_{n=1}^{N_0} w^{(k)} (n) =1,
\label{sum_w}
\end{equation}
 and the numbers $N_k$ are defined as
 % -------------------------- i ---- (7) -------
\begin{equation}
N_k= \sum_{n=1}^{N_0} n w^{(k)} (n).
\label{sumk_w}
\end{equation}
%\textcolor{blue}{
Two dimensional distribution  function  $w^{(k)} (n)$  gives complete statistical description of the system. 
However, we shall accent on the averaged distributions $N_k$,  since they are of practical interest. 
%}

% ----------------------------------------------------------------------------------------------------------------------------
% ---------------------------------------------- fig 7 7------------------------------------------------------------------------------
% ----------------------------------------------------------------------------------------------------------------------------
\begin{figure}
%\vspace{20mm}
%\includegraphics[bb=70 80 600 200, scale=0.6]{wk_n.eps}
\includegraphics[bb=50 0 620 400, width=0.55\textwidth]{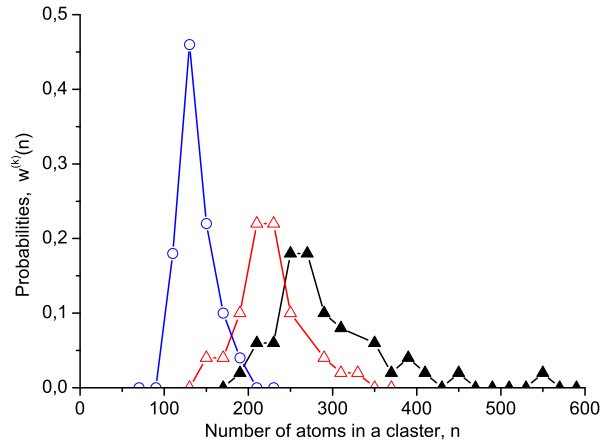}
% original eps: bb=0 0 570 400
\centering
%\vspace{10mm}
\caption{
Distribution functions $w^{(k)}(n)$  for clusters with a given number $k$ as a function of the number of particles
% in them - Deleted !!
$n$ for: $k = 1$ (black triangles, $N_1=296$); $k = 2$ (open triangles, $N_2=233$); and $k = 6$ (circles, $N_6=137$).
 }
\label{fig_wk(n)}
\end{figure}
% ---------------------------------------------- ------------------------------------------------------------------------------
% ----------------------------------------------------------------------------------------------------------------------------

Distribution functions $w^{(k)}(n)$  for $k =$ 1, 2,  6, and for  $N_0$ = 5000 
(circles in Fig. \ref{fig_a=2}) 
are shown in Fig. \ref{fig_wk(n)}. The mean values of   $N_1$ = 296,  $N_2$ = 233 and $N_6$ = 137, shown in the figure and given by expression (\ref{sumk_w}), exactly coincide with the corresponding values in Fig. \ref{fig_a=2} (curves for $N_0$ = 5000).

The curves in Fig. \ref{fig_wk(n)}  were obtained by analyzing data from 50 Monte-Carlo runs. The distribution $w^{(k)}(n)$  is wider, the larger the cluster size (i.e., the smaller $k$). As can be seen from this  figure, large clusters have a rather wide variation in the number of particles.
% contained in them. Changed 24.09.2020 !!

\section{Log-normal distribution}

Let us compare the distribution function of clusters over the number of particles
% contained in them - Deleted !! 
$W(N)$ (\ref{WN})  and obtained by computer modeling described above, with one parameter log-normal distribution
 % -------------------------- i -- lognormal distribution -- (8) -------
\begin{equation}
df(N) = \frac{1}{\sigma \sqrt{2\pi}} \exp{ \left( - \frac {1}{2\sigma^2} \ln^2 N  \right)} \frac{dN}{N},
\label{lognormal}
\end{equation}
where $\sigma$ is a dispersion  of $\ln N$. Distribution function (\ref{lognormal}) is normalized 
 % -------------------------- i ---normalization of the  lognormal distribution  - (9) -------
\begin{equation}
\int_0^\infty f(N) dN = 1 .
\label{lognornma_norml}
\end{equation}
The mean values of the number of particles in a cluster $\overline{N}$  and of its square $\overline{N^2}$ are
 % -------------------------- i ---mean and square of it for lognormal dist- (10), (11) -------
\begin{eqnarray}
\overline{N} = \int_0^\infty Nf(N) dN = \exp \left( \frac{\sigma^2}{2} \right), 
\label{lognormal_N} \\
\overline{N^2} = \int_0^\infty N^2 f(N) dN = \exp \left( 2\sigma^2  \right) . 
\label{lognormal_N2}
\end{eqnarray}
Knowing the mean number of particles in the cluster, 
%\textcolor{blue}{
$\overline{N}=N_0 /\langle k_0^{(i)} \rangle$, 
%} 
from the simulation, one can calculate the parameter $\sigma$  from Eq. (\ref{lognormal_N}) and compare the distribution (\ref{lognormal}) with results of simulation.
%\textcolor{blue}{
$\langle k_0^{(i)} \rangle$ is a mean number of clusters, averaged over 50 Monte-Carlo runs as explained in the section III. Comparison of the distribution Eq. (\ref{lognormal}) with our calculations contains, therefore, no fitting parameters.
%}

For small values of the argument in Eq. (\ref{lognormal}) $N \ll \exp (\sigma \sqrt{2})$ the log-normal distribution (\ref{lognormal}) becomes
 % -------------------------- i ---   the  lognormal distribution  for small argument - (12) -------
\begin{equation}
f(N) \approx \frac{1}{\sigma N  \sqrt{2\pi}}  \left( 1  - \frac {1}{2\sigma^2} \ln^2 N  \right).
\label{lognornmal_small}
\end{equation}

\begin{figure}
\includegraphics[bb=50 0 872 578, width=0.55\textwidth]{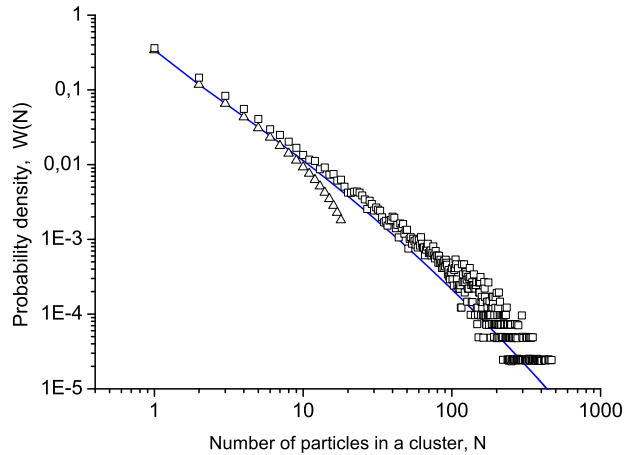}
% original eps: bb=0 0 822 578
%\vspace{10mm}
\centering
\caption{
The distribution function of clusters over the number of particles $W(N)$ for $a = 2$ and $N_0 = 10^4$ (squares); solid  line is the log-normal distribution  (\ref{lognormal});  the light triangles is the log-normal distribution for small values of the argument (\ref{lognornmal_small}). In this example, the mean number of particles in a cluster is $\overline{N}$ = 12.1  and $\sigma$ = 2.23
 }
\label{fig_log_a=2}
\end{figure}

\begin{figure}
\includegraphics[bb=80 0 902 578, width=0.55\textwidth]{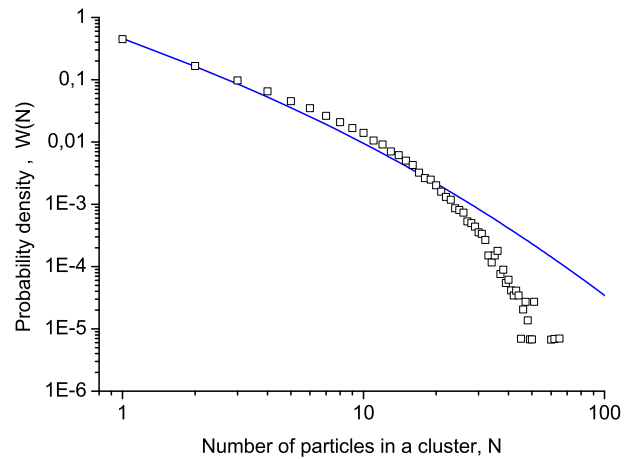}
% original eps: bb=0 0 822 578
%\vspace{10mm}
\centering
\caption{
The same as Fig. \ref{fig_log_a=2}  but for  $a = 1.5$. The log-normal distribution  (\ref{lognormal}) is shown by the full curve.  In this case  $\overline{N}$ = 3.42  and $\sigma$ = 1.57.
 }
\label{fig_log_a=1.5}
\end{figure}

Figs. \ref{fig_log_a=2} and  \ref{fig_log_a=1.5}   illustrate the comparison of the log-normal distribution   (\ref{lognormal}) with the results of the computer simulation for different values of the interaction parameter $a$ and fixed number of particles in the system $N_0 = 10^4$. It follows from these figures that log-normal distribution describes  
%satisfactory !! Changed 
%\textcolor{blue}{
the general  
%}
behaviour of the 
%\textcolor{blue}{
normalized 
%}
cluster distribution function $W(N)$  
%\textcolor{blue}{
for $a=2$, as well as for $a=1.5,$
%}  
%especially !! Deleted 
for clusters with relatively small number of particles.  One can also learn from these figures that the cluster distribution function is very sensitive to the value of the interaction parameter (\ref{a}).
%\textcolor{blue}{
Formula (\ref{lognornmal_small}) gives the correct initial slope of the  curve $W(N)$.
%}

The Monte-Carlo simulation gives a large scatter in the distribution of clusters over the number of particles $W(N)$ for big clusters (i.e., for clusters with a large number of particles). 
The behaviour of $W(N)$  is  
%\textcolor{blue}{
not regular even for each individual history (see Fig. \ref{fig_W_2_10000}). 
Also
%}
the number of particles in the largest clusters (with small $k$ in $N_k$) is different for different histories  in accordance with the distributions of $w ^ k (n) $ shown in Fig. \ref{fig_wk(n)}. This circumstances lead to a statistical straggling in the distribution $W (N)$  for large $N$, as can be seen in Figs. \ref{fig_log_a=2}  and \ref{fig_log_a=1.5}. It follows from these figures that the system contains a gas from a large number of individual atoms or small clusters described by a 
%\textcolor{blue}{
distribution similar to Eq. (\ref{lognormal})
%}
, and relatively small number of big clusters containing many particles (like Fig.7 in Ref. \cite{McDowell2013}).

%\textcolor{blue}{   
In the limit of $a \rightarrow 0$, all $N_k = 1$, $k<N_0$. In the opposite limit, $a \rightarrow \infty$, there is one cluster in the system with $N_1 = N_0$  (for example, a complete condensation of gas into liquid). Under conditions close to both limits the log-normal distribution can not describe the distribution $W(N)$. The most suitable conditions for a log-normal distribution is $a \sim 2$.
%}
% ----------------------------------------------------------------------------------------------------------------------------
% ---------------------------------------------- fig 8 8------------------------------------------------------------------------------
% ----------------------------------------------------------------------------------------------------------------------------
% ---------------------------------------------- ------------------------------------------------------------------------------
% ----------------------------------------------------------------------------------------------------------------------------
% ----------------------------------------------------------------------------------------------------------------------------
% ---------------------------------------------- fig 9 9------------------------------------------------------------------------------
% ----------------------------------------------------------------------------------------------------------------------------
%% ---------------------------------------------- fig  ------------------------------------------------------------------------------
%% ----------------------------------------------------------------------------------------------------------------------------
%\begin{figure}
%\includegraphics[bb=70 80 800 500, scale=0.4]{Nlog_1_25.eps}
%\centering
%\vspace{10mm}
%\caption{
%The same as Fig. \ref{fig_log_a=2}  but for  $a = 1.25$. The log-normal distribution  (\ref{lognormal}) is shown by the full curve.  %In this case  $\overline{N}$ = 2.05  and $\sigma$ = 1.2
 %}
%label{fig_log_a=1.25}
%\end{figure}
%% ---------------------------------------------- ------------------------------------------------------------------------------
%% ----------------------------------------------------------------------------------------------------------------------------

\section{Universal distribution}

\begin{figure}
\includegraphics[bb=80 0 902 578, width=0.55\textwidth]{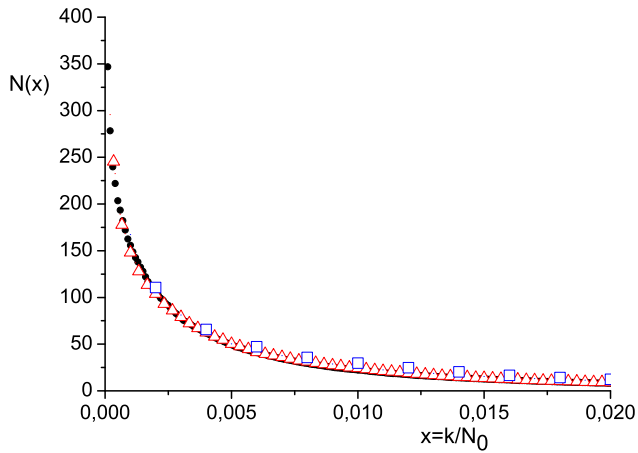}
% original eps: bb=0 0 822 578
\centering
\caption{
The distributions  of particles  over the clusters $N(x)$ as a function of the continuous variable $x=k/N_0$ (the cluster rating). Distributions correspond to the data shown in Fig. \ref{fig_a=2} for $a=2$ and different   numbers of particles in the system $N_0$:   $N_0=10^4$ (dots);   $N_0=3000$ (triangles) and  $N_0=500$ (squares).
 }
\label{fig_a=2_univ}
\end{figure}

Instead of a discrete variable $k$, representing the ranked number,  we introduce the continuous variable $x = k / N_0$. In this case, discrete distribution $N_k$ transforms  into continuous distribution  $N(x)=N_{k/N_0}$, $0<x<1$.  The distribution of particles over clusters $N(x)$, corresponding to the distributions  in Fig. \ref{fig_a=2}  for $a=2$,  are shown in Fig. \ref{fig_a=2_univ}.  For the new variable $x$, all curves $N(x)$  for different numbers  of particles in the system $N_0$  almost coincide.
The ``boundary conditions'' associated with the number of particles in the system $N_0$  play a significant role for a relatively small number of particles  $N_0 <10^3$. Therefore, it can be argued that for large $N_0>10^3$  the  function $N (x)$ represents some universal distribution of particles over clusters, depending only on one parameter which is the interaction parameter $ a$ in Eq. (\ref{a}).
The variable $x$ can be called a relative ranked cluster number, or simply a cluster rating. This name is justified by the fact that smaller values of $x$ correspond to clusters with a larger number of particles, that is, the rating of clusters with small $x$ is greater than that of clusters with larger $x$. 
%\textcolor{blue}{
Thus, $N(x)$  represents the mean number of particles in a cluster with a ranked number $k=N_0 x$ (the integer part of this product  is assumed), where $N_0$  is a number of particles in the volume of interest. 
%} 

% ---------------------------------------------- fig 10 ------------------------------------------------------------------------------
% ----------------------------------------------------------------------------------------------------------------------------
% ---------------------------------------------- ------------------------------------------------------------------------------
% ----------------------------------------------------------------------------------------------------------------------------

\begin{figure}
\includegraphics[bb=80 0 902 578, width=0.55\textwidth]{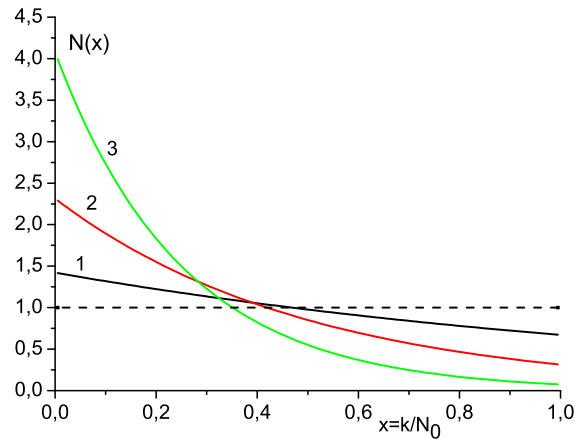}
% original eps: bb=0 0 822 578
\centering
\caption{
Illustration of the behavior of the distribution function $N(x)$  for different values of the interaction parameter $a$. The dashed line refers to the system consisting of single particles only ($a = 0$). Lines 1, 2 and 3 correspond to the increase of the parameter $a$: 
$a_3>a_2>a_1$ .
}
\label{fig_model_distrib}
\end{figure}

The distribution function $N(x)$ is normalized  
 % -------------------------- i ---   Normalization of N(x)  -------------------------------------- - (13) -------
\begin{equation}
\int_0^1 N(x)dx=1.
\label{int_N=1}
\end{equation}
In many cases, the upper limit of integration in Eq. (\ref{int_N=1}) can be extended to infinity. Schematic drawing of the behavior of the distribution  $N(x)$ as a function of the interaction parameter $a$ is shown in Fig. \ref{fig_model_distrib}. The dashed line corresponds to absence of clusters in the system ($a=0$), i.e.  the system consists of individual atoms only. As $a$  increases, the distribution function  $N(x)$ shifts to the left, so that $N(x) \rightarrow \delta (x)$ if $a \rightarrow \infty$; $\delta (x)$ is the  Dirac's $\delta$-function. The behavior of the distribution function similar to that shown in Fig. \ref{fig_model_distrib}, with the same initial and boundary conditions, takes place in the theory of cascade equations, where the role of the time variable is played by the interaction parameter $a$ (see Fig.1 in \cite{mkh2004} and related text).  A Smoluchowski-type kinetic model to  describe the evolution of the cluster size distribution in the systems of self-propelled particles has been  formulated in \cite{Peruani2010}. 
% ---------------------------------------------- fig 11 ------------------------------------------------------------------------------
% ----------------------------------------------------------------------------------------------------------------------------
% ---------------------------------------------- ------------------------------------------------------------------------------
% ----------------------------------------------------------------------------------------------------------------------------

\begin{figure}
\includegraphics[bb=80 0 902 578, width=0.55\textwidth]{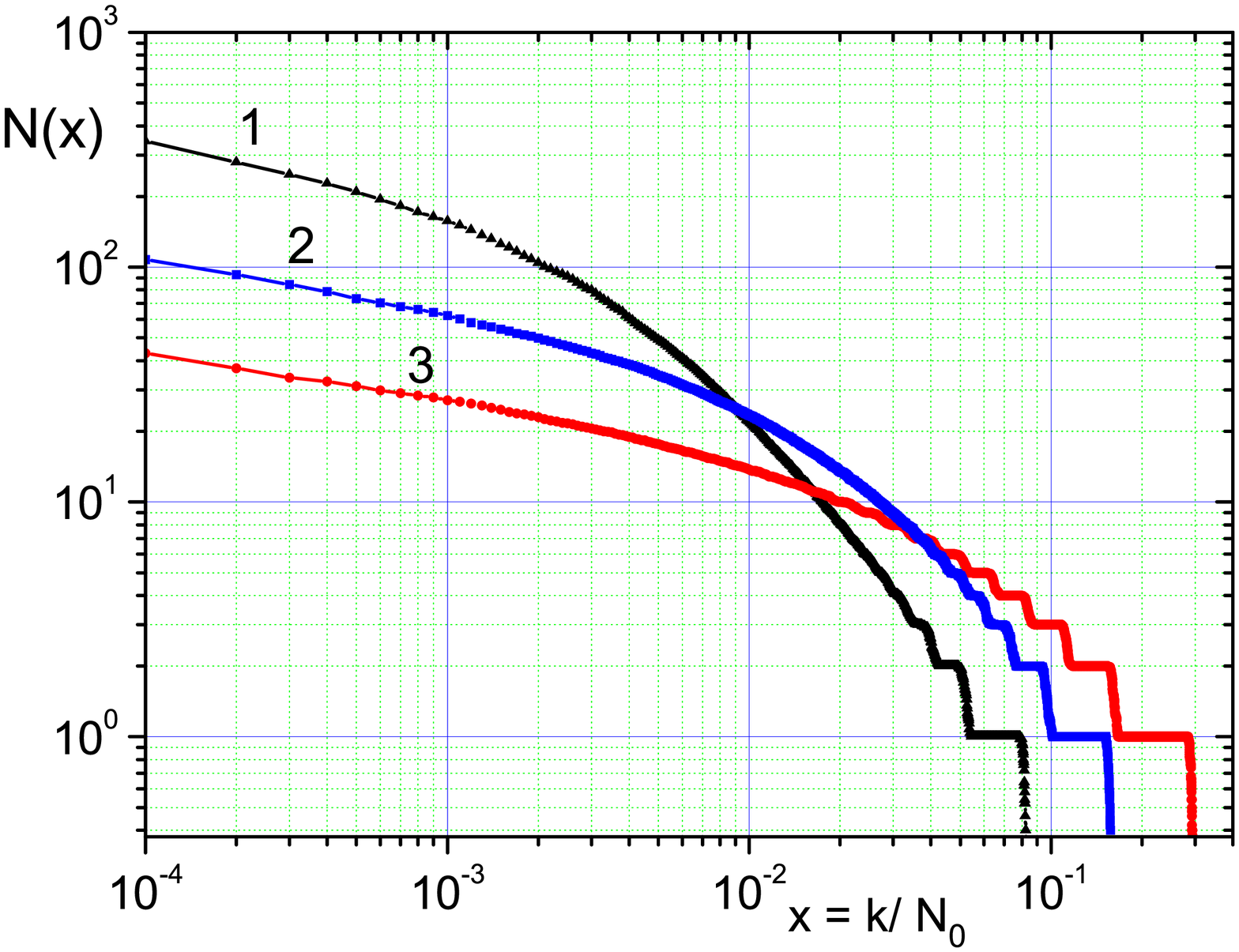}
% original eps: bb=0 0 822 578
\centering
\caption{
Universal distribution functions $N(x) $ for different values of the interaction parameter $a$: $a=2$ (curve 1), $a=1.75$ (curve 2) and $a=1.5$ (curve 3). 
}
\label{fig_lt_a=2}
\end{figure}

\begin{figure}
\includegraphics[bb=80 0 902 578, width=0.55\textwidth]{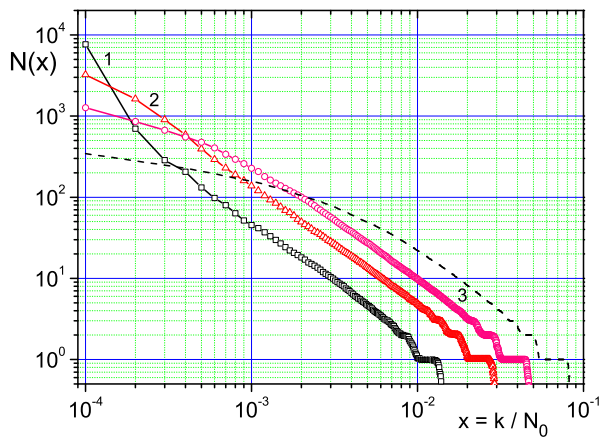}
% original eps: bb=0 0 822 578
\centering
\caption{
The same as Fig. \ref{fig_lt_a=2}  but for: $a=3$ (curve 1, squares),   $a=2.5$ (curve 2, triangles), $a=2.25$ (curve 3, circles) and $a=2$ (dashed line). 
}
\label{fig_gt_a=2}
\end{figure}

The universal distribution functions $N(x)$  for different values of the interaction parameter $a$  are shown in Figs. \ref{fig_lt_a=2} and \ref{fig_gt_a=2}. The calculations have been made for $N_0 = 10^4$. Each point is the result of averaging over 50 runs.  The area under these curves defines the fraction of particles  $\Delta N$  that are contained in clusters with a rating in the interval ($x, x+\Delta x$).
% ----------------------------------------------------------------------------------------------------------------------------
% ---------------------------------------------- fig 12 ------------------------------------------------------------------------------
% ----------------------------------------------------------------------------------------------------------------------------
% ----------------------------------------------------------------------------------------------------------------------------
% ---------------------------------------------- fig 13 ------------------------------------------------------------------------------
% ----------------------------------------------------------------------------------------------------------------------------
% ----------------------------------------------------------------------------------------------------------------------------
% ---------------------------------------------- fig 14 ------------------------------------------------------------------------------
% ----------------------------------------------------------------------------------------------------------------------------
\begin{figure}
\includegraphics[bb=80 0 902 578, width=0.55\textwidth]{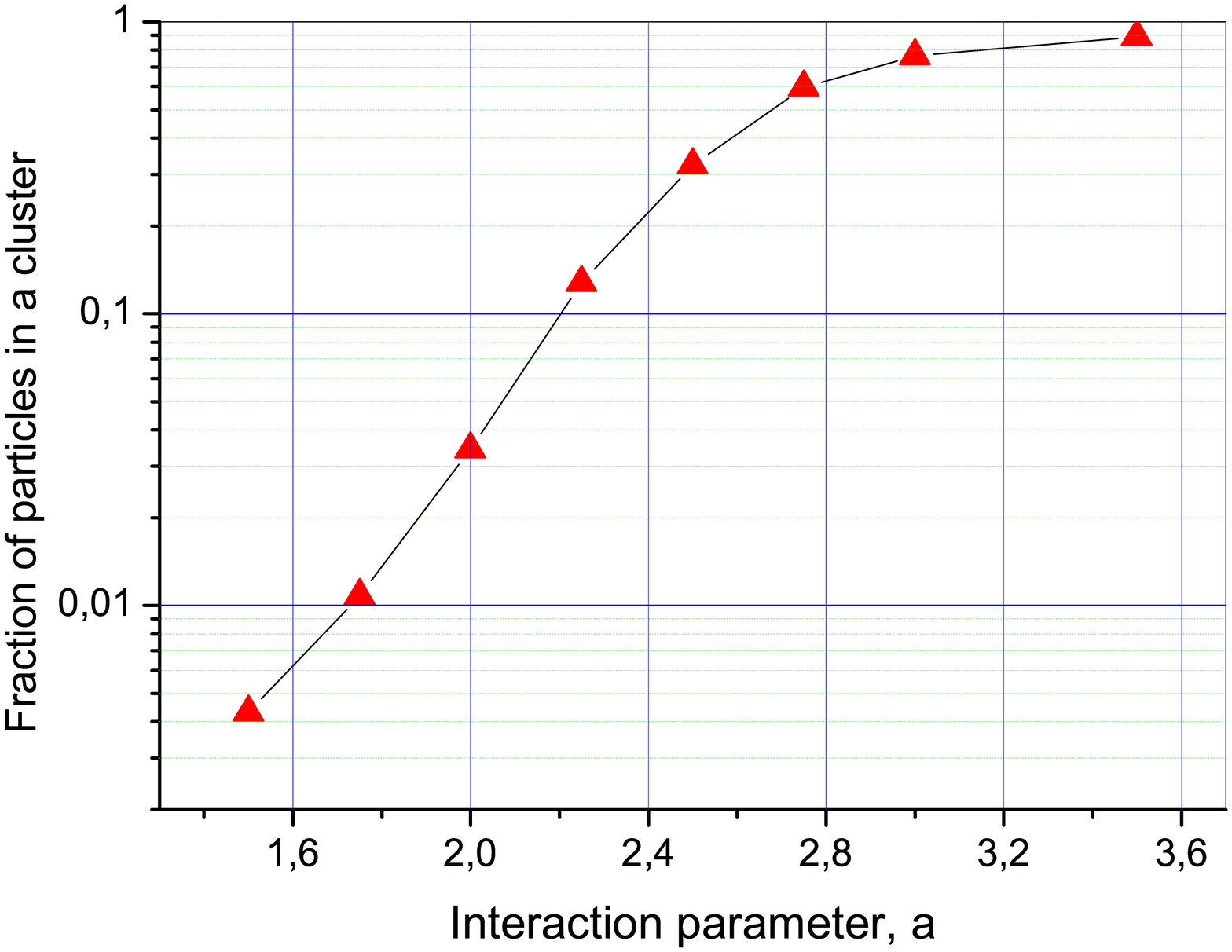}
% original eps: bb=0 0 822 578
\centering
\caption{
The fraction of particles contained in the cluster of the largest size, depending on the interaction parameter $a$.
}
\label{fig_max_cluster}
\end{figure}
% ----------------------------------------------------------------------------------------------------------------------------
% ----------------------------------------------------------------------------------------------------------------------------

The generalization of the discrete distribution of clusters over the number of particles (\ref{WN}) on the continuous distribution $W(N)dN$, which gives the probability that the cluster contains the number (or fraction) of particles in the interval $(N,N + dN)$, can be done as follows. From the dependence $N(x)$ we find the inverse function $x(N)$, after which the desired distribution has the form
 % -------------------------- i ---   cluster distribution of W(N)  -------------------------------------- - (14) -------
\begin{equation}
W(N)= \left| \frac {dx(N)}{dN} \right|,
\label{W(N)dN}
\end{equation}
where $0<x<1$,   $0<N<\infty$ and
 % -------------------------- i ---   normal of cluster distribution of W(N)  -------------------------------------- - (15) -------
\begin{equation}
\int_0^\infty W(N)dN= 1. 
\label{W(N)_norm}
\end{equation}
As an example, consider the normalized  model distribution
% -------------------------- i ---   model  distribution of N(x)  -------------------------------------- - (16) -------
\begin{equation}
N(x) = (1-\alpha)x^{-\alpha},
\label{N(x)_model}
\end{equation}
where $0<\alpha<1$.
% $N=1-\alpha$ means the uniform distribution $x=1$, so $N \ge 1-\alpha$. Then  Deleted!! 
For the distribution function of clusters over the number of particles
% in them - Deleted !! 
we find
% -------------------------- i ---   model  distribution W(N)  -------------------------------------- - (17) -------
\begin{equation}
dW(N) = \frac{(1-\alpha)^{1/\alpha}}{\alpha}  N^{ -\frac{  1+\alpha}{ \alpha}}dN.
\label{N(x)_model}
\end{equation}

In some cases, it may be useful to do the inverse procedure and reconstruct the distribution of particles among the clusters $ N (x) $ from the known distribution function $ W (N) $.

It follows from Figs. \ref{fig_lt_a=2} and \ref{fig_gt_a=2} that the process of cluster formation strongly depends on the parameter $a$ and rapidly intensifies for $a>1.5$. Thus, at $a = 1.5$, the intervals of $x$ variable for which the clusters contain 25 \%  each of all particles in the system are equal to: (0, 1.45$\cdot 10^{-2}$); (1.45$\cdot 10^{-2}$, 4.43$\cdot 10^{-2}$); (4.43$\cdot 10^{-2}$, 1.05$\cdot 10^{-1}$) and (1.05$\cdot 10^{-1}$, 3.03$\cdot 10^{-1}$).  These intervals are dramatically shifted towards smaller values of $x$.  At $ a = 2$  they are: (0, 1.2$\cdot 10^{-3}$); (1.2$\cdot 10^{-3}$, 3.9$\cdot 10^{-3}$); (3.9$\cdot 10^{-3}$, 1.13$\cdot 10^{-2}$) and (1.13$\cdot 10^{-2}$, 8.59$\cdot 10^{-2}$). When $a = 3$, already 75 \% of particles are contained in clusters with $x <10^{-4}$. The latter means, for example, that  if a system consists of $10^{4}$  particles, then there is one large cluster containing approximately $7.5\cdot 10^{3}$ particles.

This is illustrated in Fig. \ref{fig_max_cluster}, which shows the dependence of the fraction of particles contained in the cluster of the largest size on the interaction parameter $a$. When $a> 3$, almost all particles of the system are contained in one cluster, that is, complete condensation occurs. Note that this model does not take into account the fact that 
%\textcolor{blue}{
real
%}
particles cannot approach each other less than to a certain distance due to the strong repulsion of atoms at small distances. In fact, complete condensation occurs already at  $a> 2.5$  (see below).

For large values of $ x$, the stepwise character of the   curves $N (x)$  is manifested in Figs. \ref{fig_lt_a=2}, \ref{fig_gt_a=2}. The steps correspond to individual atoms, or clusters containing 2 or 3 particles. For  $N_0 \rightarrow \infty$, the stepwise behavior of the function $N (x)$ will disappear.

% ----------------------------------------------------------------------------------------------------------------------------
% ---------------------------------------------- fig 15 ------------------------------------------------------------------------------
% ----------------------------------------------------------------------------------------------------------------------------
\begin{figure}
\includegraphics[bb=80 0 902 578, width=0.55\textwidth]{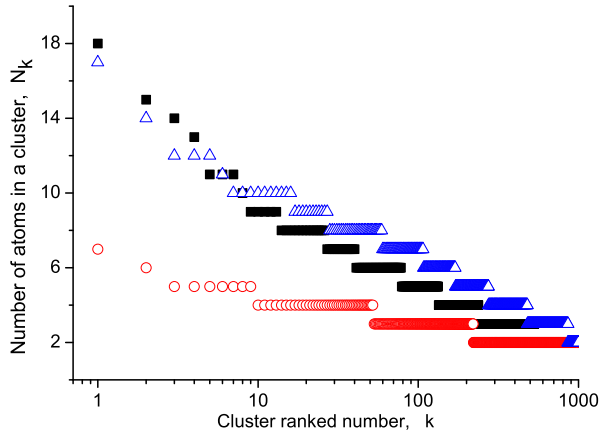}
% original eps: bb=0 0 822 578
\centering
\caption{
Distributions of argon atoms over clusters $N_k$  at a temperature of 95 K in a system with $N_0=$ 8000 atoms ($a \approx 0.8$): black square symbols -- calculation by the MD method; round symbols -- universal distribution with $a = 0.8$; triangles -- universal distribution with $a = 1.2$.
}
\label{fig_MD_95K}
\end{figure}
% ----------------------------------------------------------------------------------------------------------------------------
% ---------------------------------------------- fig 16 ------------------------------------------------------------------------------
% ----------------------------------------------------------------------------------------------------------------------------
\begin{figure}
\includegraphics[bb=80 0 902 578, width=0.55\textwidth]{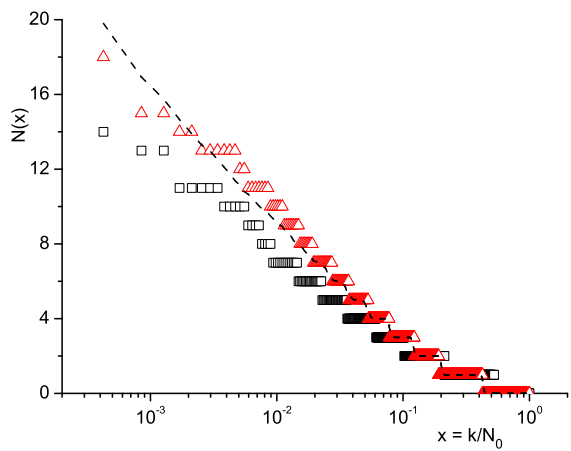}
% original eps: bb=0 0 822 578
\centering
\caption{
Distributions of argon atoms over clusters $ N_k$  in a system with $N_0=$2353 atoms  and $a =$ 1.355. Square symbols - calculation by the MD method; triangles - the universal distribution obtained with a single run; the dotted line is the universal distribution obtained over 50 runs.
}
\label{fig_MD_2353}
\end{figure}
% ----------------------------------------------------------------------------------------------------------------------------
\section{Comparison with molecular dynamics }

It should be emphasized that in this article we do not set as our goal a detailed interpretation of experimental data. 
% or real physical situations. Deleted !!
%Nevertheless, it is interesting to compare the obtained one-parameter universal distributions with the distributions in 
%\textcolor{blue}{
Our aim is to qualitatively compare the obtained distributions with calculations of real systems by the molecular dynamics method. 
%}

Further, the distributions of $ N_k $, which depend on the number of particles in the system, will also be called universal, like $ N (x) $. 
 Comparison of the distribution of atoms over clusters $N_k$ with that calculated by the molecular dynamics method (MD) for a system of 8000 krypton atoms at a temperature of 95 K is shown in Fig. \ref{fig_MD_95K}. These conditions correspond to the value of the interaction parameter  $a \approx 0.8$. The MD method used in this calculation 
is described in \cite{mkh2014, *mkh2016}, 
%\cite{mkh2016} 
and is based on LAMMPS computer package \cite{Parks2008}. 

It is seen in Fig. \ref{fig_MD_95K} that the universal distribution with $a$ = 0.8 gives a weaker degree of cluster formation for large clusters $k<$200 than that predicted by MD. This is due to the fact that the attraction of atoms takes place in real system, which contributes to the formation of clusters. In addition, we consider the static model, whereas in real systems atoms move and not all of them can form bound states in collisions. A reverse processes of destruction of bound states during collisions also occur.
For a universal distribution, this corresponds to a larger value of the interaction parameter $a$. So, in this example, at $a$ = 1.2, the model distribution quite adequately reproduces the results of molecular dynamics.

Fig. \ref{fig_MD_2353} illustrates the comparison of the MD calculation for argon at temperature $T=1.293T_0$ with the universal distribution $N_k$. $T_0$ corresponds to the minimum of the interaction  potential of argon atoms with each other. The number of atoms in the system is $N_0=2353$. Under these conditions $a\approx 1.355$. 
In this case, there is a good agreement between the universal  distribution and MD calculation.

\section{Conclusions}

A universal particle distribution function over clusters, $N(x)$, depending only on one parameter, can be introduced
%\textcolor{blue}{
for a system of overlapping spheres, which form clusters. The distribution is independent on boundary conditions and number of particles (spheres) in the system. 
%}
The parameter $a$  plays the role of the order parameter for the system under consideration and is equal to the ratio of the interaction radius to the mean distance between particles. Strictly speaking, the universal distribution refers to a system of spheres, which, at the intersection, form clusters and the centers of which are uniformly distributed in space. Nevertheless, this distribution can adequately describe realistic systems as well. 
%\textcolor{blue}{
Each point on the curve $N(x)$ shows the mean number of particles in a cluster with a rank number 
$ k=N_0 x$, where $N_0$  is a number of particles in the volume of interest, and  $0<x<1$.
The rank $k$ is assigned to clusters according to the cluster sizes. The biggest cluster has rank $k=1$.   
%}

%\textcolor{blue}{
The universal curve $N(x)$ is obtained  from the discrete distribution, $N_k$, which gives the mean number of particles in a cluster with a rank number $k$,   in the limit, $N_0 \rightarrow \infty$.   
The function $N_k$  is in unique relation with conventionally used distribution $W(N)$, which shows the probability that cluster contains exactly $N$ particles. In contrast with $N_k$, the big clusters correspond to the tail of the distribution $W(N)$. If we are interested in processes of cluster formation, all interesting things happen on this tail. From this point of view the distribution $N_k$  is more convenient, since big clusters correspond to the values of $N_k$ close to the maximum, i.e. small rank numbers $k$.  
%}
The universal distribution 
%\textcolor{blue}{
reduces to the behaviour similar to
%}
a log-normal distribution for
%\textcolor{blue}{
$a \sim 2$.
%}
%but differs significantly from it for larger values of $a$, the proportion of which may be small, but which %contain a large number of particles, and, therefore, define the basic features of the system. Deleted!!

There are three main factors that lead to a difference in the distributions of particles over clusters in real systems from the universal distributions considered in this paper. The main factor is that atoms in real systems cannot approach unboundedly close to each other. This leads to the fact that already for $a>2$, clusters in real systems already contain more particles than the universal distribution predicts. The second factor is that the universal distributions suggest that if two atoms are located close 
%\textcolor{blue}{
enough 
%}
to each other, then they inevitably form a bound state. In fact, it is not. This factor can be taken into account if 
to consider only those moving particles that can form bound states energetically. I.e. $N_0$ in this case should be the number of those particles for which the kinetic energy in their center of mass is less than the absolute value of the  interaction potential barrier depth. Finally, the third factor is the presence of an attractive force between atoms, which leads to more intense cluster formation, especially under conditions close to the conditions of condensation.

This work was supported by the grants of Russian Foundation for Basic Research No. 18-02-01042 A and the Fund for the Promotion of Innovation (grant No. 0038507).

\section*{Appendix}
 In what follows we shall consider the algorithm for calculating the cluster structure and illustrate it for the  system shown 
in Fig. \ref{fig_N=21}.  

\begin{figure}
\includegraphics[bb=80 0 904 575, width=0.55\textwidth]{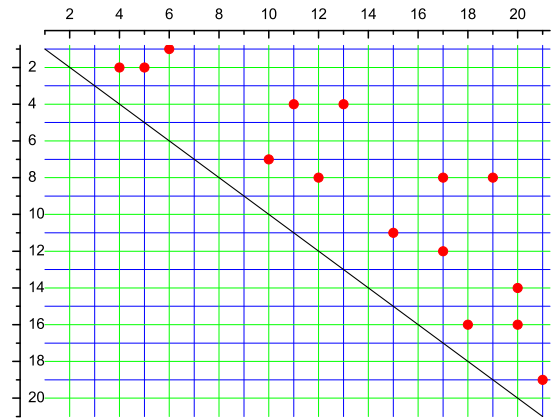}
% original eps: bb=0 0 824 575
\centering
\caption{
The matrix  $Q$  for the system of 21 particles shown in Fig. 1. Each point corresponds to a binary bond between particles, i.e. to the element of type ($a_k, b_k$)  in Eq. (B). 
}
\label{cluster_matrix}
\end{figure}

Let the matrix  $R_{ij}$  define the distances between the particles ($i<j$ ). We construct a square matrix  $Q$  with zero diagonal elements and zeros for all elements below the main diagonal. The remaining elements (above the diagonal) consist of zeros and ones, so that
% ----------------------------------- ---------------------------- App Eq. (A) ---------------------
$$
Q_{ij}= \left\{
\begin{array}{rcl}
0, & \mbox{if} &  R_{ij}>R; \\
1, & \mbox{if} &  R_{ij} \le R;
\end{array}
\right.
\eqno(\mbox{A})
$$
% ------------------------------------------------------------------------------------------------------
where $R$ is the ``interaction radius''; $(i,j)=1,...,N_0$; $N_0$ is the number of particles in the system. The number of ones  in such a matrix is equal to the number of binary bonds in the system. The matrix $Q$  corresponding to the system in 
Fig.\ref{fig_N=21}  is shown in Fig.\ref{cluster_matrix}. Instead of ones, the elements of the matrix are shown by bold points. The matrix in this case contains 15 points, corresponding to 15 bonds between particles.

A system without bonds (i.e., without clusters) corresponds to a zero matrix. Further, the indices of the rows and columns $i,j$  of the matrix for some  $k$-th bond will be denoted as  $(a_k,b_k)$, where $a_k$  and $b_k$   are integers, numbering the rows and columns of the matrix  $Q$.

% ----------------------------------------------------------------------------------------------------------------------------
% ---------------------------------------------- fig 17 matrix--------------------------------------------------------------
% ----------------------------------------------------------------------------------------------------------------------------
% ----------------------------------------------------------------------------------------------------------------------------

The set of   $m$  links (bonds)
% ---------------------------------------------------------------------------- App. (B) --------------
$$
(a_1,b_1),(a_2,b_2),...,(a_m,b_m),
\eqno(\mbox{B})
$$
% ------------------------------------------------------------------------------------------------------
will belong to one cluster, if one of the numbers $a_i$   or $b_i$   (or both) of any pair occurs at least once in the remaining pairs in (B). For example, a cluster of six particles in Fig.\ref{fig_N=21} corresponds to a set of five bonds 
%(see the second matrix row in Fig. \ref{cluster_matrix}) 
% ---------------------------------------------------------------------------- App. (C) --------------
$$
(2,4),\, (2,5), \, (4,11),\, (4,13),\, (11,15).
\eqno(\mbox{C})
$$
% ------------------------------------------------------------------------------------------------------
We see that at least one of the numbers in any pair is contained at least once in one of the other pairs. From the construction (C) we conclude that this cluster consists of 6 particles with numbers: 2, 4, 5, 11, 13 and 15. The rows and columns of the matrix $ Q$, which correspond to the configuration (C), form a grid that defines the particles that belong to the same cluster. 

Collections of pairs of type (B), and hence the distribution of particles in clusters, can be found by analyzing the rows of the matrix $Q$, starting with the first row. All non-zero elements of a row or column of the matrix $Q$  belong to the same cluster. The bonds with the same row or column numbers belong to the same cluster as well (for example, when we consider another row with the same number as the column number in the original matrix element, see below). If all elements of the  $k$-th row and  $k$-th column are equal to zero, then the $k$-th particle does not form bonds and is a cluster of one  particle.

Analysis of the cluster structure of the matrix $Q$  begins with the 1st row, which defines all the bonds of the first particle with the other particles (for a given system configuration, the particles are numbered in an arbitrary order). All these particles are included in the first cluster. Further, we do not mean the ranked cluster number. Clusters will be numbered in the order of their formation. The ranked number will be determined after identifying all the clusters in the system followed by lining up in order of decreasing the number of particles in them.
%and only then line up in order of decreasing the number of particles in them.- Changed !! 

If a bond (i.e., a unit matrix element) appears in the 1st row and  $k$-th column, then the particles with numbers 1 and $k$  belong to the same cluster. Similarly, the remaining bonds in the first line $(1,j)$,  $j>k$, are revealed. After that, we go to the line (matrix row) with the number $k$  and reveal all the bonds in this line. The particles forming these bonds are also included into the cluster 1. Then, it is analyzed whether there are links in the remaining rows with numbers $j$  that have already been encountered as column numbers in the 1st row. In subsequent calculations, all rows and columns belonging to the same cluster are no longer considered.

After all the particles belonging to cluster 1 formed by the first particle (i.e., the 1st matrix row) are identified, then we go to the second matrix row. If the number 2 has already entered the sequence of links (B) included in the first cluster, then we go to the 3rd matrix row, otherwise we analyze the matrix row 2, as described above. And so on, we iterate through the rows of the matrix $Q$. 

As an example, consider how the sequence of bonds (C) is formed. This sequence is formed by the second row of the matrix  $Q$  in Fig.\ref{cluster_matrix}. There are the unit elements (2, 4) and (2, 5) on this line. Therefore, we go first to the 4th row. It forms bonds (4, 11) and (4, 13). Lines 5 and 13 do not form bonds, but line 11 contains the link (11, 15). This gives the sequence (C). Further, the rows and columns with numbers included in the sequence (C), fall out of the analysis, and we go to the matrix line 3, etc.

As a result, with the help of the matrix $Q$  in Fig.\ref{cluster_matrix}, we arrive at the following sequence of calculation of the cluster structure of the system shown in Fig.\ref{fig_N=21}.

Cluster 1: (1,6), 2 particles with numbers 1 and 6;

Cluster 2: is defined by the sequence (C);

Cluster 3: one particle with number 3;

Cluster 4: (7,10), 2 particles with numbers 7 and 10;

Cluster 5: (8,12), (8,17), (8,19), (12,17), (19, 21), 5 particles with numbers 8, 12, 17, 19, 21;

Cluster 6: one particle with number 9;

Cluster 7: (14,20), (16,18), (16,20), 4 particles with numbers 14, 16, 18, 20.

All matrix lines after the 14th fall out of the analysis, since all particles are already distributed in clusters.

%\newpage

%\input{lit_cluster.tex}

	\nocite{*}

	\bibliography{cluster_pr3}

%\begin{thebibliography}{33}

%\end{thebibliography}

\end{document}